\documentstyle[12pt,epsf]{article}

\begin{document}
\renewcommand{\thefootnote}{\fnsymbol{footnote}}
\renewcommand{\theenumi}{(\roman{enumi})}
\renewcommand{\thefigure}{\arabic{figure}}
\title{Exact Parametrization \\of The Mass Matrices and The KM Matrix}
\author{{K. Harayama}\thanks{E-mail address  harayama@eken.phys.nagoya-u.ac.jp} \mbox{\quad} and \mbox{\quad} {N. Okamura}\thanks{E-mail address  okamura@eken.phys.nagoya-u.ac.jp}\\ \\{\it Department of Physics, Nagoya University,}\\{\it Nagoya 464-01, Japan}\\ {\it Telephone:81(52)789-2450 \quad Fax:81(52)789-2860}}
\date{ }
\maketitle
\vspace{-11cm}
\begin{flushright}
DPNU-96-24\\
hep-ph/9605215\\
May 1996
\end{flushright}
\vspace{7cm}
\begin{abstract}
We analyze properties of general quark mass matrices.
The up and down part quark mass matrices are written in terms of
six dimensionless parameters and six quark masses.
It is shown that two of the former six dimensionless parameters
can be chosen to be any value.
Once values for these two parameters are chosen,
Kobayashi-Maskawa matrix is written
in terms of the remaining four parameters.
Our results are given analytically without any approximation.
\end{abstract}
\section{Introduction}
\par
\hspace*{\parindent}
The standard model (SM) explains current high energy experiments,
but it offers no understanding of many parameters included in the SM
(e.g. the fermion masses and the flavor mixing angles).
It is expected that there exist some fundamental theory which includes
the SM as a low energy effective theory.
This theory should offer a deeper understanding of these parameters.
If some reasonable relations among the SM parameters
could be discovered phenomenologically, it would provide
us with important clue toward the search for the fundamental theory.

Recently the top quark has been discovered and its mass has been measured
\cite{{Top1},{Top2}}.
More precise data on elements of the Kobayashi-Maskawa (KM) Matrix
\cite{KM}
will be available in the near future, e.g. from the experiments at
the $B$-factories.
Thus it is timely to study the quark mass matrices and
the KM matrix through various phenomenological approaches.
Up to now, many attempt have been made to
investigate the explicit connection between the quark mixing matrix and
quark mass matrices \cite{Fritzsch}-\cite{Raby}.
Here we present exact relations between the general mass matrices and
the flavor mixing matrix.
This approach
is obviously advantageous over many specific $ans{\ddot{a}}tze$ and
can model-independently shed light on the origin of quark masses
and flavor mixing.

In this work, we start from the nearest-neighbor interactions (NNI) form
of quark mass matrices.
Branco {\it et al.} have shown that any $3\times3$ quark mass matrices
can be transformed into the NNI basis form both up and down part
at the same time, so we do not lose any generality by using
this basis \cite{NNI}.
The up and down mass matrices have twelve parameters,
we fix six of them so that the six quark masses can be correctly reproduced.
Further more we show that two real parameters can be chosen 
to be {\it any} value.
The remaining four parameters are obtained from the measurements of the
KM matrix.

This paper is organized as follows.
In section {\ref{sec:NNI}},
we review the NNI form of the mass matrices.
In section {\ref{sec:free}},
we separate out the two arbitrary degrees of freedom of the NNI basis.
These two degrees of freedom can be chosen implicitly and will not change any
observable.
In section {\ref{sec:epsf}},
we show how to separate the elements of quark mass matrices 
in terms of quark masses and dimensionless parameters on the NNI basis.
In section {\ref{sec:exact_KM}}, 
exact analytical calculation of the KM matrix is presented.
In section {\ref{sec:sum}},
we specialize our result to the Fritzsch $ans{\ddot{a}}tze$
and summarize our results.

\section{The Nearest Neighbor Interactions (NNI) Form}
\label{sec:NNI}
{\it NNI Basis}
\par
It has been shown that the arbitrary up and down
part $3\times3$ quark mass matrices can be simultaneously transformed into
the NNI form
without changing any observable quantities \cite{NNI}.
That is, this transformation itself does not change the values
of quark masses and the KM matrix elements.
Hence the NNI form includes all physical contents of quark mass
matrices in the SM.
In this basis each mass matrix has four texture zeros,
so it is very simple.
Generally, $3\times3$ mass matrices in the NNI basis
can be written as follows:
\begin{eqnarray}
M_{u} &=& m_{t} M_{u}^{\prime} = m_{t} \left( \begin{array}{ccc}
                                                0     & a_{u} & 0     \\
                                                c_{u} & 0     & b_{u} \\
                                                0     & d_{u} & e_{u}
                                            \end{array} \right) ,
\nonumber \\
M_{d} &=& m_{b} M_{d}^{\prime} = m_{b} \left( \begin{array}{ccc}
                                                0     & a_{d} & 0     \\
                                                c_{d} & 0     & b_{d} \\
                                                0     & d_{d} & e_{d}
                                            \end{array} \right) ,
\label{eqn:defm}
\end{eqnarray}
where $m_{t}$ and $m_{b}$ are the top and bottom quark masses, respectively.
Note that in the NNI form, $a_{u} \sim e_{u}$ and $a_{d} \sim e_{d}$ 
are chosen real and non-negative and the phases are moved to a 
diagonal phase matrix $P$ shown below.
$M_{u} M_{u}^{T}$ and $M_{d} M_{d}^{T}$ are diagonalized by
orthogonal matrices $O_{u}$ and $O_{d}$ through
\begin{eqnarray}
O_{u}^{T} M_{u} M_{u}^{T} O_{u}
         = \left( \begin{array}{ccc}
             m_{u}^{2} & 0         & 0         \\
             0         & m_{c}^{2} & 0         \\
             0         & 0         & m_{t}^{2}
           \end{array} \right) ,
\nonumber \\
O_{d}^{T} M_{d} M_{d}^{T} O_{d}
         = \left( \begin{array}{ccc}
             m_{d}^{2} & 0         & 0         \\
             0         & m_{s}^{2} & 0         \\
             0         & 0         & m_{b}^{2}
           \end{array} \right) ,
\label{eqn:OMMO}
\end{eqnarray}
where $m_{u}$, $m_{c}$, $m_{d}$ and $m_{s}$ are
the up, charm, down and strange quark masses, respectively.
We can write the KM matrix $V_{KM}$ by reparametrizing the right handed
quark field phases,
\begin{equation}
V_{KM} = O_{u}^{T} P O_{d} ,
\end{equation}
where
\begin{equation}
P = \left( \begin{array}{ccc}
              1 & 0                               & 0                 \\
              0 & \exp\left( i \theta_{2} \right) & 0                 \\
              0 & 0                 & \exp\left( i \theta_{3} \right)
           \end{array} \right)
\label{eqn:Phase}
\end{equation}
is the phase-difference matrix between the up and down quark sectors
\cite{{KM},{Fritzsch},{IT}}.

\section{Remaining Degrees of Freedom}
\label{sec:free}
{\it Degrees of Freedom}
\par
There are twelve real parameters in the above NNI form,
although the number of observable parameters is ten,
that is, the six quark masses and four measurements of the KM matrix.
Remaining two degrees of freedom are those of the choice of the NNI basis.

Any mass matrices $\hat{M_{u}}=m_{t}\hat{M_{u}}^{\prime}$
and $\hat{M_{d}}=m_{b}\hat{M_{d}}^{\prime}$ can be transformed
into the above NNI basis mass matrices $M_{u}=m_{t}M_{u}^{\prime}$
and $P M_{d}=m_{b} P M_{d}^{\prime}$ without changing the values of
the quark masses and the KM matrix through
\begin{equation}
U^{\dagger} \hat{M_{u}}^{\prime} V_{u} = M_{u}^{\prime}, \mbox{\quad}
U^{\dagger} \hat{M_{d}}^{\prime} V_{d} = P M_{d}^{\prime},
\label{eqn:defN}
\end{equation}
where $V_{u}$, $V_{d}$ and $U$ are unitary matrices.
(Here, the matrix $P$ is displayed in (\ref{eqn:Phase}).)
In order to construct $U$, we choose some non-vanishing complex
value $k$ at first.
Then, using $k$, we obtain $U_{i2}$ $(i=1\sim3)$
as the eigenvector of the matrix
$ \hat{M_{u}}^{\prime} \hat{M_{u}}^{\prime \dagger} + k \hat{M_{d}}^{\prime} \hat{M_{d}}^{\prime \dagger} $
as follows:
\begin{equation}
 \left( \hat{M_{u}}^{\prime} \hat{M_{u}}^{\prime \dagger}
    + k \hat{M_{d}}^{\prime} \hat{M_{d}}^{\prime \dagger}
 \right)_{ji} U_{i2}
= \lambda U_{j2},
\label{eqn:Mu+Md}
\end{equation}
where $\lambda$ is the eigenvalue of the matrix
$ \hat{M_{u}}^{\prime} \hat{M_{u}}^{\prime \dagger} + k \hat{M_{d}}^{\prime} \hat{M_{d}}^{\prime \dagger} $ and $i,j=1\sim3$.
So $k$ is arbitrary degrees of freedom of the NNI basis.
Using equations (\ref{eqn:defN}) and (\ref{eqn:Mu+Md}),
we can write $k$ and $\lambda$
in terms of the elements of the NNI basis mass matrices
$M_{u}=m_{t}M_{u}^{\prime}$ and $P M_{d}=m_{b} P M_{d}^{\prime}$
as follows:
\begin{equation}
k
= - \frac{b_{u}e_{u}}{b_{d}e_{d}}
      \exp\left\{i\left(\theta_2-\theta_3\right)\right\}, \mbox{\quad}
\lambda= b_{u}^{2}+c_{u}^{2} + k \left( b_{d}^{2}+c_{d}^{2} \right).
\label{eqn:k}
\end{equation}
Therefore we can fix two arbitrary degrees of freedom of
the NNI basis by fixing the complex number $k$.
Obviously from the above calculations, this fixing does not
require any constraint to the quark masses and the KM matrix.

In another way, we can construct $U$ by starting from $U_{i1}$ $(i=1\sim3)$
as follows.
$U_{i1}$ is given as the eigenvector of the matrix
$ \hat{M_{u}}^{\prime} \hat{M_{u}}^{\prime \dagger} + k^{\prime} \hat{M_{d}}^{\prime} \hat{M_{d}}^{\prime \dagger} $.
Here, $k^{\prime}$ is some arbitrary complex and non-vanishing value.
Then, instead of (\ref{eqn:Mu+Md}), we use following equation:
\begin{equation}
 \left(          \hat{M_{u}}^{\prime} \hat{M_{u}}^{\prime \dagger}
    + k^{\prime} \hat{M_{d}}^{\prime} \hat{M_{d}}^{\prime \dagger}
 \right)_{ji} U_{i1}
= \lambda^{\prime} U_{j1},
\label{eqn:Mu+Md2}
\end{equation}
where $\lambda^{\prime}$ is the eigenvalue of the matrix
$ \hat{M_{u}}^{\prime} \hat{M_{u}}^{\prime \dagger} + k^{\prime} \hat{M_{d}}^{\prime} \hat{M_{d}}^{\prime \dagger} $ and $i,j=1\sim3$.
In this way, ${k^{\prime}}$ is also arbitrary degrees of freedom.
Using equations (\ref{eqn:defN}) and (\ref{eqn:Mu+Md2}),
we obtain the expressions for $k^{\prime}$ and $\lambda^{\prime}$
as follows:
\begin{equation}
k^{\prime}
= - \frac{a_{u}d_{u}}{a_{d}d_{d}}\exp\left( -i \theta_{3} \right), \mbox{\quad}
\lambda^{\prime}= a_{u}^{2} + k^{\prime} a_{d}^{2}.
\end{equation}
Also, Fixing the value of $k^{\prime}$, instead of fixing the value of $k$,
does not require any constraint to the KM matrix and the quark masses.

If we fix the arbitrary two degrees of freedom of basis, $k$ or $k^{\prime}$,
then ten degrees of freedom remain in the NNI basis mass matrices.

\section{Quark Masses and Dimensionless Parameters}
\label{sec:epsf}
\par
\hspace*{\parindent}
In this section we present the parametrization of the quark mass matrices
on the NNI basis in terms of six eigenvalues
and six dimensionless parameters without any approximation.
\\
\\
{\it Parameter Separating}
\par
The characteristic equation of the matrix
\(M^{\prime} M^{\prime T}\) is
\begin{eqnarray}
 &-&\xi^{3} + \left(A+B+C+D+E\right)\xi^{2} \nonumber \\
 & & - \left(AB+AC+AE+BD+CD+CE\right)\xi + ACE = 0,
\label{eqn:char}
\end{eqnarray}
where $A=a^{2}$, $B=b^{2}$, $C=c^{2}$, $D=d^{2}$ and $E=e^{2}$, and
$\xi$ is an eigenvalue of \(M^{\prime} M^{\prime T}\).
Because the textures of up and down mass matrices are the same,
we discuss them in a common manner by omitting the indexes $u$ and $d$.
We know that $\xi$ in the equation (\ref{eqn:char}) has three solutions:
\begin{eqnarray}
\xi &=& \xi_{1} = \frac{m_{1}^{2}}{m_{3}^{2}},
\nonumber \\
\xi &=& \xi_{2} = \frac{m_{2}^{2}}{m_{3}^{2}},
\nonumber \\
\xi &=& \xi_{3} = 1,
\label{eqn:EVa}
\end{eqnarray}
where $m_{i}$ are the quark masses of the $i$-th generation.
Obviously $\xi$ are dimensionless,
because $M_{u}^{\prime}$ and $M_{d}^{\prime}$
have been normalized by the third generation quark masses.
Therefore,
\begin{eqnarray}
A+B+C+D+E &=& \xi_{1} + \xi_{2} + 1, \\
\label{eqn:a+b+}
AB+AC+AE+BD+CD+CE &=& \xi_{1} + \xi_{2} +  \xi_{1} \xi_{2}, \\
\label{eqn:ab+ac+}
ACE &=& \xi_{1} \xi_{2}.
\label{rqn:ace}
\end{eqnarray}
We introduce two real and positive mass parameters as follows:
\begin{eqnarray}
p &=& \xi_{1} + \xi_{2},
\nonumber \\
q^{4} &=& \xi_{1} \xi_{2}.
\end{eqnarray}
Note that the equation (\ref{rqn:ace}) is satisfied by introducing
two real and positive dimensionless parameters, $y$ and $z$:
\begin{eqnarray}
A &=& \frac{ q^{2} z^{2} }{ y^{2} }, \nonumber \\
C &=& \frac{ q^{2} }{ y^{2} z^{2} }, \nonumber \\
E &=& y^{4}.
\label{eqn:deface}
\end{eqnarray}
Although $y$ and $z$ are independent of the quark masses,
they enter the KM matrix.
Furthermore, equation (\ref{eqn:a+b+}) and equation (\ref{eqn:ab+ac+})
can be written as
\begin{eqnarray}
\left(B+C\right)+\left(D+A\right)&=& p + 1 - E,\nonumber \\
\left(B+C\right)\left(D+A\right) &=& p + q^{4} - E\left(A+C\right).
\label{eqn:def2}
\end{eqnarray}
Therefore, $B+C$ and $D+A$ are solutions of
the following quadratic equation of $\zeta$:
\begin{equation}
\zeta^{2} - \left( p + 1 - E \right) \zeta + p + q^{4} - E\left(A+C\right) = 0.
\label{eqn:eqn_zeta}
\end{equation}
Solving equation (\ref{eqn:eqn_zeta}) analytically, we obtain
\begin{eqnarray}
B &=& \frac{1}{2}\left\{p+1-y^{4} \pm \sqrt{\left(1-p+y^{4}\right)^{2}-4\left(q^{2}-y^{2} z^{2}\right)\left(q^{2}-\frac{ y^{2} }{ z^{2} }\right)}\right\} - \frac{ q^{2} }{ y^{2} z^{2} } , \nonumber \\
D &=& \frac{1}{2}\left\{p+1-y^{4} \mp \sqrt{\left(1-p+y^{4}\right)^{2}-4\left(q^{2}-y^{2} z^{2}\right)\left(q^{2}-\frac{ y^{2} }{ z^{2} }\right)}\right\} - \frac{ q^{2} z^{2} }{ y^{2} } .
\label{eqn:defbd}
\end{eqnarray}
As a consequence, we can write general quark mass matrices
in the following simple form without any approximation:
\begin{equation}
M^{\prime}\\
= \left( \begin{array}{ccc}
            0                 & \displaystyle\frac{ q z }{ y } & 0         \\
            \displaystyle\frac{ q }{ y z } & 0                 & \sqrt{ B }\\
            0                 & \sqrt{ D }        & y^{2}
         \end{array} \right) .
\label{eqn:case}
\end{equation}
Note that two possibilities of signs are included in equation
(\ref{eqn:defbd}),
so let us call
    the case that the plus-minus sign in $B$ ($D$) is plus (minus) case (I)
and the case that the plus-minus sign in $B$ ($D$) is minus (plus) case (II).
\\
\\
{\it Mathematically Allowed Regions for $y$ and $z$}
\par
Next, we discuss mathematically allowed regions for $y$ and $z$.
Each element of the mass matrices can be taken to be non-negative values
through redefinition of the quark field phases,
as discussed at the beginning.
Equation (\ref{eqn:deface}) means that
$A>0$, $C>0$ and $E>0$.
That is, \( \det{M} \neq 0 \).
About $B$ and $D$, they must satisfy the constraints
$B \geq 0$ and $D \geq 0$.
This implies that values in the square roots in equation (\ref{eqn:defbd})
are non-negative.

The mathematically allowed regions for $y$ and $z$ with the non-negativity
constraints are illustrated in Fig.1(a),(b) and Fig.2(a),(b).

We point out that $B$ and $D$ do not vanish at the same time
due to the following restrictions: the CP-violating phase should exist
and no element of the KM matrix should be zero.

\section{The KM matrix}
\label{sec:exact_KM}
\par
\hspace*{\parindent}
We can write down the KM matrix without any approximation
by using above form.
\\
\\
{\it Orthogonal Matrices}
\par
When both $B$ and $D$ are not zero,
eigenvectors of $M^{\prime}$$M^{\prime T}$
can be written in terms of eigenvalues as follows:
\begin{equation}
\left (
\begin{array}{r}
\alpha_i \\
\beta_i \\
\gamma_i
\end{array}
\right )
=
\frac{1}{f_i}
\left (
\begin{array}{c}
(\xi_i-B-C)ad \\
(\xi_i-A)be \\
(\xi_i-A)(\xi_i-B-C)
\end{array}
\right ) ,
\label{eqn:EVe}
\end{equation}
where $\xi_i$ are eigenvalues of \(M^{\prime} M^{\prime T}\)
given in (\ref{eqn:EVa}) and
$f_i$ are normalization factors of the eigenvectors.
The orthogonal matrix in equations ({\ref{eqn:OMMO}})
can be written in terms of the above eigenvectors as follows:
\begin{eqnarray}
O &=&
\left(
\begin{array}{ccc}
\alpha_1 & \alpha_2 & \alpha_3 \\
\beta_1  & \beta_2  & \beta_3  \\
\gamma_1 & \gamma_2 & \gamma_3
\end{array}
\right) \nonumber \\
&=&
\left(
\begin{array}{ccc}
\displaystyle \frac{(\xi_1-B-C)ad}{f_1}        &\displaystyle \frac{(\xi_2-B-C)ad}{f_2}        &\displaystyle \frac{(\xi_3-B-C)ad}{f_3} \\
\displaystyle \frac{(\xi_1-A)be}{f_1}          &\displaystyle \frac{(\xi_2-A)be}{f_2}          &\displaystyle \frac{(\xi_3-A)be}{f_3}          \\
\displaystyle \frac{(\xi_1-A)(\xi_1-B-C)}{f_1} &\displaystyle \frac{(\xi_2-A)(\xi_2-B-C)}{f_2} &\displaystyle \frac{(\xi_3-A)(\xi_3-B-C)}{f_3}
\end{array}
\right) .
\nonumber \\
\label{eqn:OthE}
\end{eqnarray}
\\
\\
{\it Normalization Factors}
\par
Normalization factors $f_{i}$ can be written as follows:
\begin{eqnarray}
f_i^2 &=& AD(\xi_i-B-C)^2 \nonumber \\
      & &+BE(\xi_i-A)^2 \nonumber \\
      & &+(\xi_i-A)^2(\xi_i-B-C)^2.
\label{eqn:NF1}
\end{eqnarray}
({\ref{eqn:NF1}}) is the fourth-order in $\xi_i$.
Using the equation ({\ref{eqn:char}}),
({\ref{eqn:NF1}}) becomes the second-order in $\xi_i$.
We introduce new variables $S$, $P$ and $Q$
which can be written in terms of quark masses as follows:
\begin{eqnarray}
S &=& \xi_{1} + \xi_{2} + 1 = 1 + p, \nonumber \\
P^2 &=& \xi_{1}\xi_{2} = q^{4}, \nonumber \\
Q &=& \xi_{1}\xi_{2} + \xi_{1} + \xi_{2} = p + q^{4}.
\label{eqn:DefSPQ}
\end{eqnarray}
Normalization factors ({\ref{eqn:NF1}}) can be written
by using $Y=y^2$, $Z=z^2$ and above variables
$S$, $P$ and $Q$
\begin{eqnarray}
f_i^2 &=& \left\{ -2Q + \frac{PS{Z}}{2Y} - \frac{3PYZ}{2} + \frac{S{Y}^2}{2} +\frac{S^2}{2} \right. \nonumber \\
      & & \left. \pm \left(\frac{3PZ}{Y}-S\right)\frac{\sqrt{R}}{2}\right\}{\xi_i}^2 \nonumber \\
      &+& \left\{ 3P^2 - QY^2 - \frac{PS^2Z}{Y} + \frac{2PQZ}{Y} + PSYZ\right. \nonumber \\
      & & \left.\pm \left(Q - \frac{PSZ}{Y}\right)\sqrt{R}\right\}{\xi_i} \nonumber \\
      &+& \left\{ \frac{PQSZ}{2Y} - \frac{P^2S}{2} -\frac{PQYZ}{2} + \frac{3P^2Y^2}{2} - \frac{3P^3Z}{Y}\right. \nonumber \\
      & & \left. \pm \frac{1}{2}\left(\frac{PQZ}{Y}-3P^2\right)\sqrt{R}\right\},
\label{eqn:NF3}
\end{eqnarray}
where $R = \left(S-{Y}^2\right)^2+4P{Y}\left( {Z}+\displaystyle\frac{1}{{Z}} \right)-4Q$.
\\
\\
{\it Elements of the KM Matrix}
\par
The elements of the KM matrix become as follows:
\begin{eqnarray}
(V_{KM})_{ij} &=& \frac{1}{f_{ui}f_{dj}}\left\{ \alpha_{ui}\alpha_{dj}
                + \beta_{ui}\beta_{dj} \exp\left( i \theta_2 \right)
                + \gamma_{ui}\gamma_{dj} \exp\left(i\theta_3\right) \right\}
\nonumber \\
             &=& \frac{1}{f_{ui}f_{dj}}
\left[
       \left\{\xi_{ui}-\frac{1}{2}\left(S_u-Y_u^2\pm\sqrt{R_u} \right) \right\}
       \sqrt{
                      \frac{P_uZ_u}{2Y_u}\left(S_u-Y_u^2
                         -\frac{2P_uZ_u}{Y_u}\mp\sqrt{R_u}\right)
                      }
\right .
\nonumber \\
     & &\times
       \left\{\xi_{di}-\frac{1}{2}\left(S_d-Y_d^2\pm\sqrt{R_d} \right) \right\}
       \sqrt{
                       \frac{P_dZ_d}{2Y_d}\left(S_d-Y_d^2
                          -\frac{2P_dZ_d}{Y_d}\mp\sqrt{R_d}\right)
            } \nonumber \\
            &+&
       \left(\xi_{ui}-\frac{P_uZ_u}{Y_u}\right)
       \sqrt{
             \frac{Y_u^2}{2}\left(S_u-Y_u^2
            -\frac{2P_u}{Y_uZ_u}\pm\sqrt{R_u}\right)
            }
\nonumber \\
      & &\times
       \left(\xi_{di}-\frac{P_dZ_d}{Y_d}\right)
       \sqrt{
             \frac{Y_d^2}{2}\left(S_d-Y_d^2
            -\frac{2P_d}{Y_dZ_d}\pm\sqrt{R_d}\right)
            }
      \exp\left(i\theta_2 \right) \nonumber \\
           &+&
      \left\{\xi_{ui}-\frac{1}{2}\left(S_u-Y_u^2\pm\sqrt{R_u} \right) \right\}
      \left(\xi_{ui}-\frac{P_uZ_u}{Y_u}\right)\nonumber \\
& &\times
\left.
      \left\{\xi_{di}-\frac{1}{2}\left(S_d-Y_d^2\pm\sqrt{R_d} \right) \right\}
      \left(\xi_{di}-\frac{P_dZ_d}{Y_d}\right)
      \exp\left( i\theta_3 \right)
\right] .
\label{eqn:DKM2}
\end{eqnarray}
The expressions of elements of the KM matrix (\ref{eqn:DKM2})
is written by eigenvalues of the mass matrices, the parameters
$z_u$, $y_u$, $z_d$, and $y_d$ included in up and down part
mass matrices, and two phases.
Indexes $u$ and $d$ show the up and down part, respectively,
in equation (\ref{eqn:DKM2}).
It is written in terms of only the eigenvalues of the mass matrices and
the other six dimensionless parameters.
\\
\\
{\it More Texture Zeros Case}
\par
Equation (\ref{eqn:DKM2}) is satisfied
only when elements $a$, $b$, $c$, $d$ and $e$ of up and down part 
mass matrices are positive definite.
Though $a_{u,d}$, $c_{u,d}$ and $e_{u,d}$ are not zero,
there are cases that $b_{u,d}$ or $d_{u,d}$ $= 0$.
However, two values of $b_{u}$, $d_{u}$, $b_{d}$ and $d_{d}$ do not
become zero at the same time by physical conditions
that CP-violating phase should exist
and that no element of the KM matrix should be zero.
Therefore, we restrict our discussion to the case where one of
$b_u$, $b_d$, $d_u$ and $d_d$ is zero.

At first, let us discuss about the case of $b_{u}=0$.
In this case, up and down part mass matrices are
\begin{eqnarray}
M_{u} &=& m_{t} M_{u}^{\prime} = m_{t} \left( \begin{array}{ccc}
                                                0     & a_{u} & 0     \\
                                                c_{u} & 0     & 0     \\
                                                0     & d_{u} & e_{u}
                                            \end{array} \right) ,
\nonumber \\
M_{d} &=& m_{b} M_{d}^{\prime} = m_{b} \left( \begin{array}{ccc}
                                                0     & a_{d} & 0     \\
                                                c_{d} & 0     & b_{d} \\
                                                0     & d_{d} & e_{d}
                                            \end{array} \right) ,
\label{eqn:defSM}
\end{eqnarray}
except phases.
One of the eigenvalues of $M_{u}^{\prime}$ $M_{u}^{\prime T}$ is $C_{u}$.
The eigenvector associated with the eigenvalue $C_{u}$ is
\begin{equation}
\left(
\begin{array}{c}
0 \\
1 \\
0
\end{array}
\right).
\label{eqn:EVe_C}
\end{equation}
The other two eigenvectors related to eigenvalues ${\xi_{u}}_{i}$ are
\begin{equation}
\frac{1} {\left( {f_{u}^{B}} \right)_{i}}
\left(
\begin{array}{c}
a_u d_u \\
0 \\
{\xi_{u}}_i-A_u
\end{array}
\right),
\label{eqn:EVe_O}
\end{equation}
where $\left({f_{u}^{B}}\right)_{i}$ are normalization factors:
\begin{equation}
\left({f_{u}^{B}}\right)_{i}^{2} = (E_{u}+D_{u}-A_{u}){\xi_{u}}_i - A_{u}(E_{u}-D_{u}-A_{u}).
\label{eqn:NFUB}
\end{equation}
Here, we have used the characteristic equation of
$M_{u}^{\prime}$ $M_{u}^{\prime T}$.
Therefore, the KM matrix can be written as follows:
\begin{eqnarray}
& &C_{u} = {\xi_{u}}_{k} \mbox{\quad(k-th eigenvalue)}, \nonumber \\
(V_{KM})_{ij} &=&
\left\{
\begin{array}{ll}
{\beta_{d}}_{j}
               & (i=k),\\
\displaystyle \frac{1}{\left({f_{u}^{B}}\right)_{i}}
       \left\{a_{u}d_{u} {\alpha_{d}}_{j}
           + ( {\xi_{u}}_{i} - A_{u}) {\gamma_{d}}_{j}
              \exp\left(i\theta_3\right)
        \right\}
               & (i \neq k),
\end{array}
\label{eqn:KMUB}
\right.
\end{eqnarray}
where $\alpha_{dj}$, $\beta_{dj}$ and $\gamma_{dj}$ are
displayed in (\ref{eqn:EVe}).

In the same way, we can write down the KM matrix
in the case of $d_{u}=0$, $b_{d}=0$ and $d_{d}=0$ as follows.
\\
\\
Case of $d_{u}=0$:
\begin{eqnarray}
& &A_{u} = {\xi_{u}}_{k} \mbox{\quad(k-th eigenvalue)}, \nonumber \\
(V_{KM})_{ij} &=&
\left \{
\begin{array}{ll}
{\alpha_d}_j
            & ( i=k ),\\
\displaystyle \frac{1}{\left(f_{u}^{D}\right)_{i}}
         \left[\left({\xi_{u}}_{i} -E_u \right) {\beta_{d}}_j
              + b_{u}e_{u} {\gamma_d}_j
               \exp\left\{i\left(\theta_3-\theta_2\right)\right\}
         \right]
             &(i \neq k),
\end{array}
\right. \nonumber \\
\left(f_{u}^{D}\right)_{i}^{2} &=& (C_u+B_u-E_u){\xi_u}_i - E_u(C_u-B_u-E_u).
\label{eqn:KMUD}
\end{eqnarray}
\\
\\
Case of $b_{d}=0$:
\begin{eqnarray}
& &C_d = {\xi_d}_k\mbox{\quad(k-th eigenvalue)}, \nonumber \\
(V_{KM})_{ik} &=&
\left\{
\begin{array}{ll}
{\beta_u}_i & (j=k), \\
\displaystyle \frac{1}{\left(f_{d}^{B}\right)_{j}}
        \left\{a_dd_d {\alpha_u}_i
               + \left({\xi_d}_j - A_d\right) {\gamma_u}_i
                \exp\left(i \theta_3\right)
        \right\}
&(j \neq k),
\end{array}
\right. \nonumber \\
\left(f_{d}^{B}\right)^{2}_j &=& (E_d+D_d-A_d){\xi_d}_j - A_d(E_d-D_d-A_d).
\label{eqn:KMDB}
\end{eqnarray}
\\
\\
Case of $d_{d}=0$:
\begin{eqnarray}
& &A_d = {\xi_d}_k \mbox{\quad(k-th eigenvalue)},
\nonumber \\
(V_{KM})_{ik} &=&
\left \{
\begin{array}{ll}
{\alpha_u}_i & (j=k), \\
\displaystyle \frac{1}{\left(f_{d}^D\right)_{j}}
         \left[ \left({\xi_d}_j - E_d \right) {\beta_u}_i
          + b_de_d {\gamma_u}_i
          \exp\left\{i\left(\theta_3-\theta_2\right)\right\}
         \right]
& (j \neq k),
\end{array}
\right. \nonumber \\
\left(f_{d}^{D}\right)_{j}^{2} &=& (C_d+B_d-E_d){\xi_d}_j - E_d(C_d-B_d-E_d).
\label{eqn:KMDD}
\end{eqnarray}

\section{Summary and Discussion}
\label{sec:sum}
\hspace*{\parindent}
{\it Fritzsch Texture}\par
The Fritzsch-type mass matrices \cite{Fritzsch} are a special case
of the NNI form.
They can be reproduced by taking $a=c$ and $b=d$
in the above form.
Explicitly, if the conditions
\begin{eqnarray}
y &=& \sqrt { 1 - \sqrt{ p - 2 q^{2} } } = \sqrt{1+\frac{m_1}{m_3}-\frac{m_2}{m_3}}, \nonumber \\
z &=& 1
\label{eqn:Fritzschyz}
\end{eqnarray}
are satisfied, then
$M_{u}$ and $M_{d}$ become the Fritzsch-type mass matrices.
\\
\\
{\it Summary}\par
Starting from general mass matrices in the NNI form, 
we presented a parametrization which 
guarantees the six eigenvalues (quark masses). 
It is written in terms of six parameters. 
We show that two of these parameters can be 
chosen implicitly. 
These two parameters change neither 
the eigenvalues nor the KM matrix elements. 
We then presented an analytic form of the KM matrix elements
in terms of these parameters. 
Our formulation is very useful for generating a class of 
$ans{\ddot{a}}tze$ for the KM matrix elements.
\\
\\
\begin{center}
{\bf {\LARGE Acknowledgments}}
\end{center}

We drew inspiration of this work form lectures at '95 Ontake summer school.

We would like to thank Prof. A. I. Sanda and Dr. Z. Z. Xing their
reading the manuscript and giving some constructive suggestions.
 We are also grateful to Dr. T. Ito for useful discussions.

\newpage
\begin{center}
{\bf{\LARGE Figure Captions}}\\
\end{center}
Fig. 1(a): Mathematically Allowed Region
for $y_u$ and $z_u$ for the case (I).
We use following values of the quark mass ratios:
\[
\frac{m_u}{m_t} = 1.60\times10^{-5}, {\mbox{\quad}} \frac{m_c}{m_t} = 3.77\times10^{-3}.
\]
\\
Fig. 1(b): Mathematically Allowed Region
for $y_u$ and $z_u$ for the case (II).
We use the same values of the quark mass ratios as those in Fig. 1(a).
\\
\\
Fig. 2(a): Mathematically Allowed Region
for $y_d$ and $z_d$ for the case (I).
We use following values of the quark mass ratios:
\[
\frac{m_d}{m_b} = 1.67\times10^{-3}, {\mbox{\quad}} \frac{m_s}{m_b} = 3.35\times10^{-2}.
\]
\\
Fig. 2(b): Mathematically Allowed Region
for $y_d$ and $z_d$ for the case (II).
We use the same values of the quark mass ratios as those in Fig. 2(a).
\newpage
\begin{minipage}[h]{\textwidth}
\vspace{-2.5cm}
\special{epsfile = 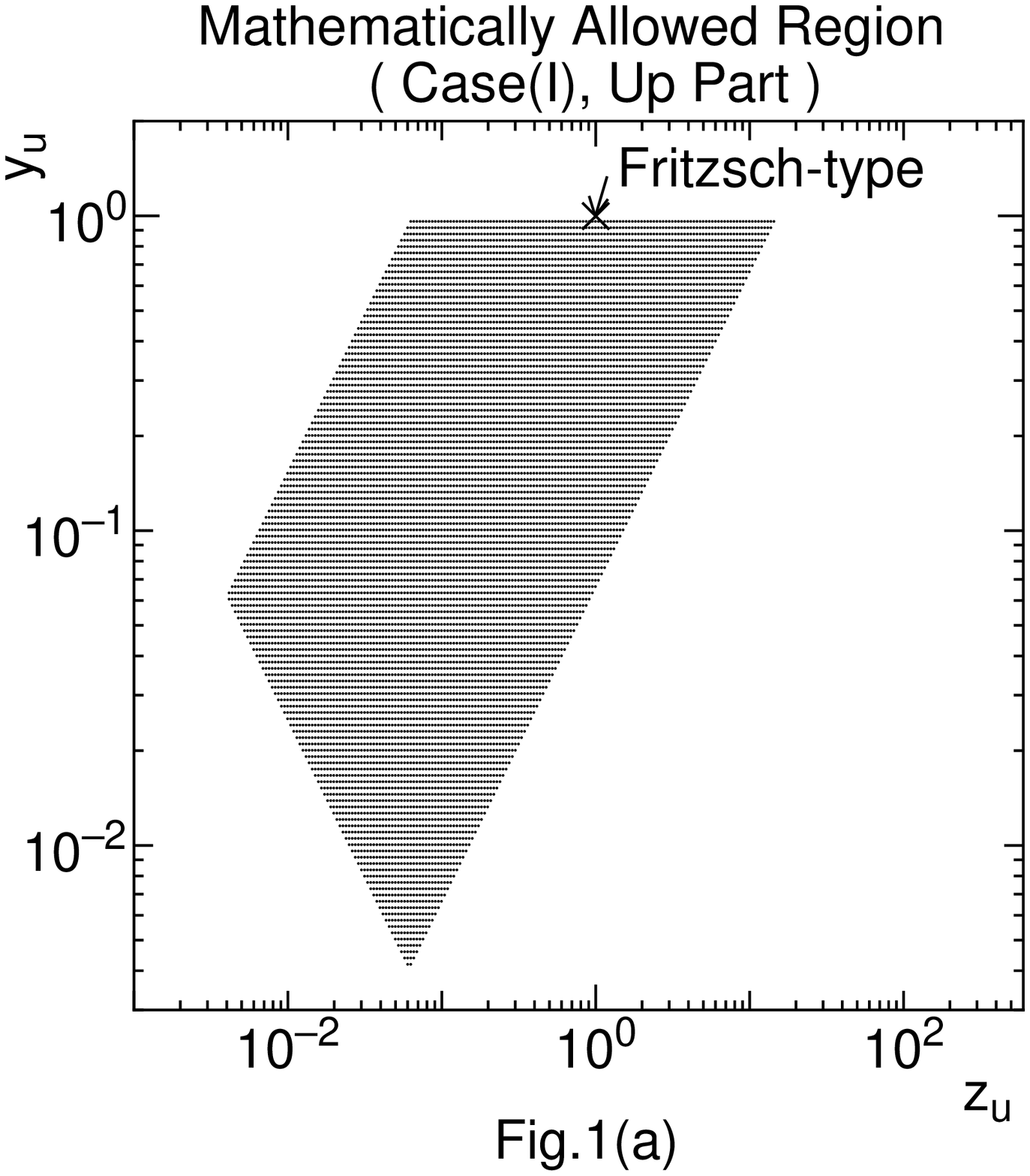
         hscale=0.7 vscale =0.7 
        }
\vspace{13cm}
\end{minipage}

\begin{minipage}[h]{\textwidth}
\special{epsfile = 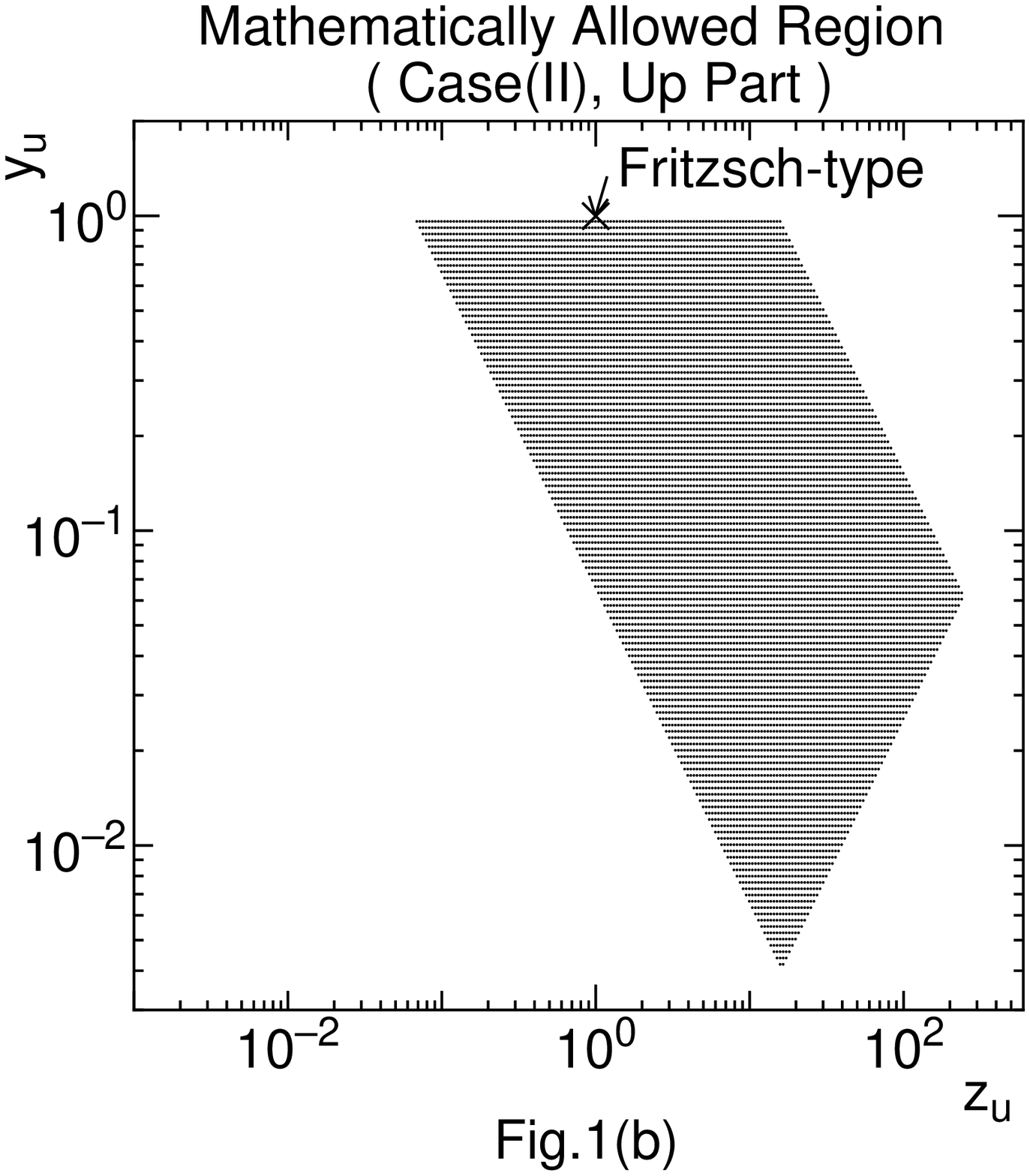
         hscale=0.7 vscale =0.7 
        }
\end{minipage}
\newpage
\begin{minipage}[h]{\textwidth}
\vspace{-2.5cm}
\special{epsfile = 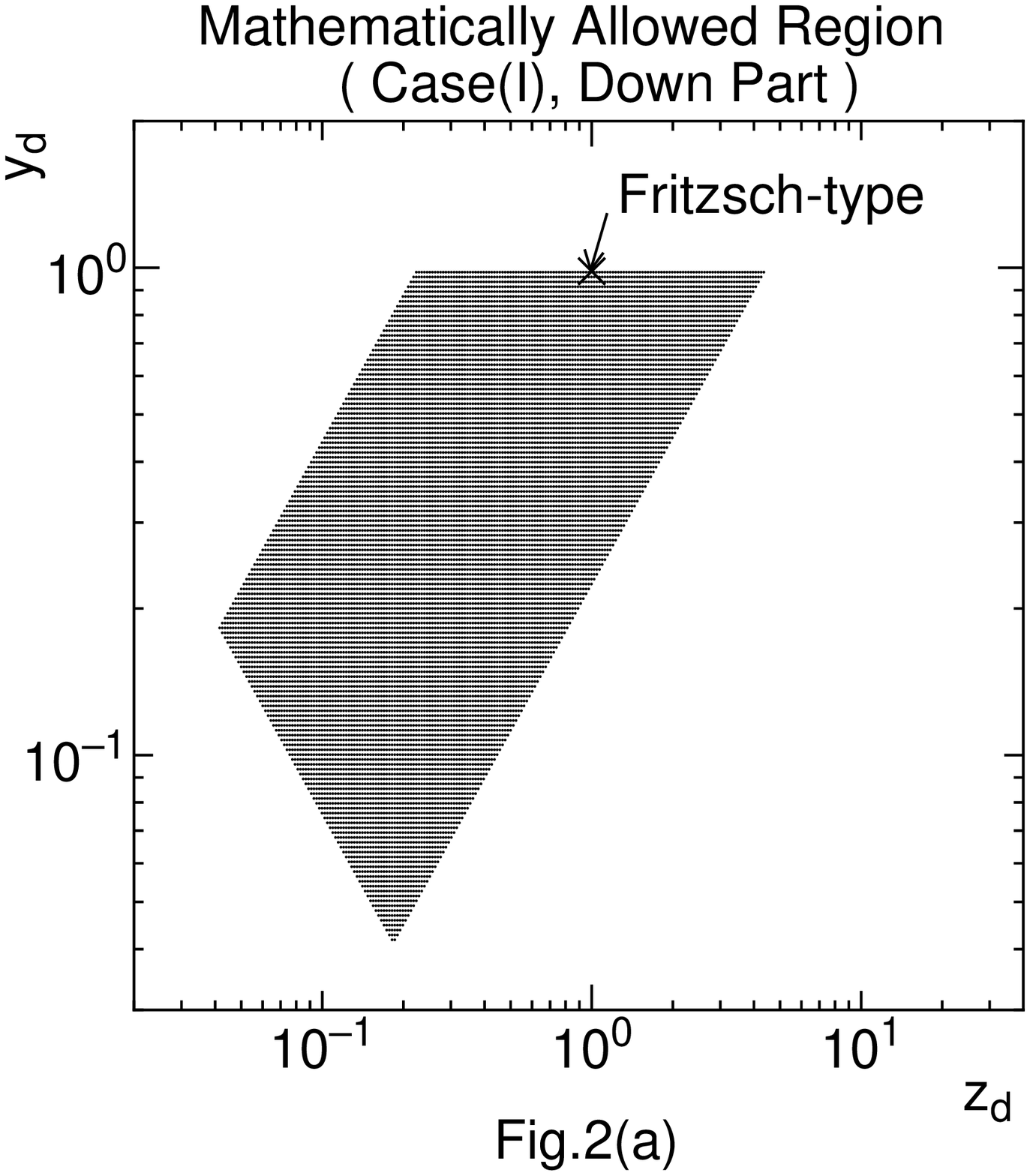
         hscale=0.7 vscale =0.7 
        }
\vspace{13cm}
\end{minipage}

\begin{minipage}[h]{\textwidth}
\special{epsfile = 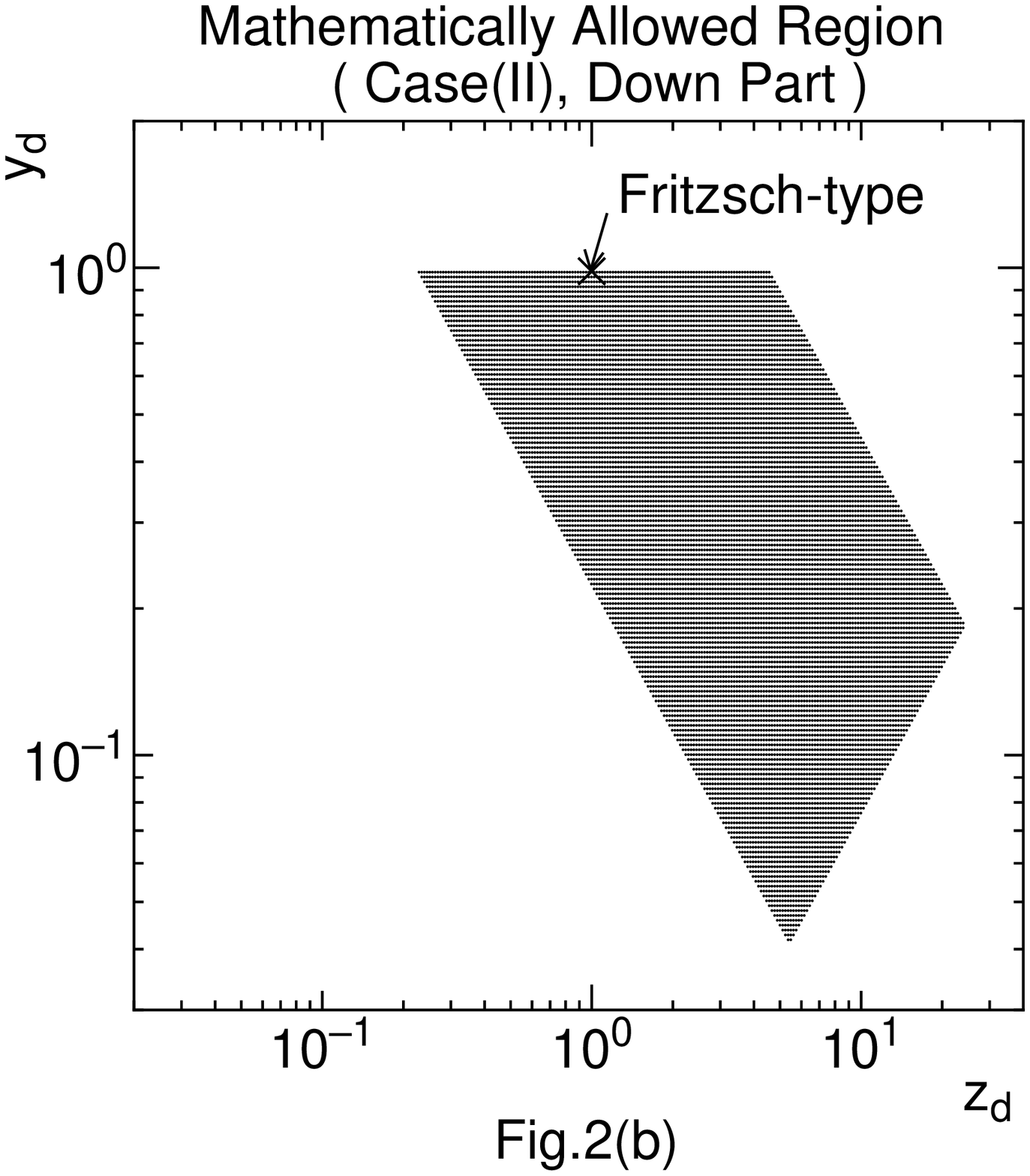
         hscale=0.7 vscale =0.7 
        }
\end{minipage}
\end{document}